\documentclass[a4paper,11pt]{article}
\usepackage{pos}

\usepackage{physics}
\usepackage{simplewick}

\newcommand\beq{ \begin{eqnarray} }
\newcommand\eeq{ \end{eqnarray} }

\usepackage{color}

\title{Measurement of hadron masses in 2-color finite density QCD}
\ShortTitle{Measurement of hadron masses in 2-color finite density QCD}

\author*[a,b]{Kotaro Murakami}
\author[c,d]{Daiki Suenaga}
\author[e]{Kei Iida}
\author[b,d,f]{Etsuko Itou}

\affiliation[a]{\it {Center for Gravitational Physics, 
Yukawa Institute for Theoretical Physics, Kyoto University, Kitashirakawa Oiwakecho, Sakyo-ku, 
Kyoto 606-8502, Japan}}
\affiliation[b]{\it {Interdisciplinary Theoretical and Mathematical Sciences Program (iTHEMS), RIKEN, Wako 351-0198, Japan}}
\affiliation[c]{\it {Strangeness Nuclear Physics Laboratory, RIKEN Nishina Center, Wako 351-0198, Japan}}
\affiliation[d]{\it {Research Center for Nuclear Physics (RCNP), Osaka University, Osaka 567-0047, Japan}}
\affiliation[e]{{\it
Department of Mathematics and Physics, Kochi University, 2-5-1 Akebono-cho, Kochi 780-8520, Japan}}
\affiliation[f]{\it {Department of Physics, and Research and 
Education Center for Natural Sciences, Keio University, 4-1-1 Hiyoshi, Yokohama, Kanagawa 223-8521, Japan}}

\emailAdd{kotaro.murakami@yukawa.kyoto-u.ac.jp}

\abstract{
We investigate hadron spectra in 2-color QCD using lattice simulation with $N_{f}=2$ at low temperature and finite density in which there appears not only the hadronic phase but also the superfluid phase. 
We first calculate the pion and rho meson spectrum, which is well-known from previous works. 
The spectral ordering of these mesons flips around the quark chemical potential $\mu=m^{0}_{\pi}/2$ ($m^{0}_{\pi}$: the pion mass at $\mu=0$), where the phase transition between the hadronic and superfluid phases occurs. 
For $\mu \gtrsim m^{0}_{\pi}/2$, the effective mass for the pion linearly increases while the one for the rho meson monotonically decreases.
Furthermore, we measure hadron spectra with the isospin $I=0$ and the angular momentum $J^{P}=0^{\pm}$.
The effective masses for the meson, diquark, and antidiquark with the same quantum number become degenerate just below $\mu = m^{0}_{\pi}/2$, and the three hadrons have the same mass in the superfluid phase.
It suggests that mixing occurs between spectra associating with mesons and baryons due to the $U(1)_{B}$ symmetry breaking.
This phenomenon can be explained in the linear sigma model with the approximate $SU(4)$ Pauli-Gursey symmetry. 
}

\FullConference{%
The 39th International Symposium on Lattice Field Theory,\\
8th-13th August, 2022,\\
Rheinische Friedrich-Wilhelms-Universit\"{a}t Bonn, Bonn, Germany
}

\begin{document}
\raisebox{80pt}[0pt][0pt]{\hspace*{92mm} YITP-22-146, RIKEN-iTHEMS-Report-22}
\vspace{-9mm}
\maketitle

\section{Introduction}
Exploration of Quantum Chromodynamics (QCD) in dense matter is one of the biggest issues. 
In particular, it is important to investigate the microscopic structure such as hadron spectrum, which underlies the macroscopic properties of dense QCD. 
At zero quark chemical potential ($\mu=0$), the lightest hadron is the pion ($I=1, J^P=0^-$ meson), a property that can be proven if one neglects the disconnected diagrams and assumes the $\gamma_5$ hermicity. At finite $\mu$ where the $\gamma_5$ hermicity is broken, the spectral ordering of hadrons becomes nontrivial.
Several analytical studies of the hadron spectrum in dense matter have been carried out based on, e.g., hadron effective models~\cite{Brown:1991kk} and QCD sum rules~\cite{Hatsuda:1991ez}. 
Also, for any number of color QCD, the $\mu$ dependence of the hadron masses can be shown to follow
\beq
m(\mu)=m(0)-Q_{B}\mu,\label{eq:hadron-linear-mass}
\eeq
with the baryon charge $Q_{B}$, in the hadronic phase~\cite{Murakami}.
In lattice study, however, such a $\mu$ dependence of the hadron spectrum is extremely hard to see due to the sign problem at low temperature. 

In this work, we consider dense 2-color QCD theory, in which lattice simulations are available even at finite density thanks to the pseudo-reality of quark fields under the $SU(2)$ gauge symmetry. 
Several lattice studies exploring the phase structure in 2-color QCD have been made (see references in Ref.~\cite{Iida:2022hyy}); at low temperature, the hadronic phase turns into the superfluid phase at $\mu\approx m^{0}_{\pi}/2$ ($m^{0}_{\pi}$: the pion mass at $\mu=0$), where the diquark condensate $\langle qq\rangle$ becomes nonzero. 

We attempt to investigate the $\mu$ dependence of the hadron spectrum in 2-color QCD with $N_{f}=2$ in a low-temperature and finite-density regime. 
The previous works have observed nontrivial features in the superfluid phase: the spectral ordering of the pion and rho meson~\cite{Muroya2002-qc, Hands2007-vp} and the existence of the (pseudo) Nambu-Goldstone mode associated with the diquark condensation~\cite{Hands2007-vp, Wilhelm:2019fvp}.
In this paper, we focus on the hadron masses with the isospin $I=0$ and the total angular momentum $J^{P}=0^{\pm}$ as well as the pion and rho meson masses.

\section{Formulation}
\subsection{Lattice action}
For $2$-color lattice QCD action, we utilize the Iwasaki gauge and the two-flavor Wilson fermion actions.  As in the case of Ref.~\cite{Iida:2019rah},
we also add the quark number operator and the diquark source term to the above fermion action as 
\beq
S_F= \bar{\psi}_1 \Delta(\mu)\psi_1 + \bar{\psi}_2 \Delta(\mu) \psi_2 - J \bar{\psi}_1 K \bar{\psi}_2^{T} + J \psi_2^T K \psi_1.\label{eq:action}
\eeq
Here, the indices $1$, $2$ denote the flavor label, 
and $K=(C \gamma_5) \tau_2$ where $C$ is the charge conjugation matrix and $\tau_2$ is the Pauli matrix acting on color indices. The additional parameter $J=j\kappa$ denotes the diquark source parameter, where $j$ and $\kappa$ denote the source parameter in the corresponding continuum theory and the hopping parameter, respectively. Here, we assume that $J$ takes a real
value. The Wilson-Dirac operator including the number operator, $\Delta(\mu)$, is defined by
\beq 
\Delta(\mu)_{x,y} = \delta_{x,y} 
&-& \kappa \sum_{i=1}^3  \left[ ( 1 - \gamma_i)  U_{x,i}\delta_{x+\hat{i},y} + (1+\gamma_i)  U^\dagger_{y,i}\delta_{x-\hat{i},y}  \right] \nonumber\\ 
&-& \kappa   \left[ e^{+\mu}( 1 - \gamma_4)  U_{x,4}\delta_{x+\hat{4},y} + e^{-\mu}(1+\gamma_4)  U^\dagger_{y,4}\delta_{x-\hat{4},y}  \right].
\eeq

\subsection{Fermion propagators}
The fermion propagators in this action are expressed as follows:
\beq \label{eq:props_wrt_cont}
\contraction[1ex]{}{\psi}{_f(x)}{\bar{\psi} }
\psi_f(x) \bar{\psi}_f(y) &=&  Q^{-1}(\mu) \Delta^{\dag}(\mu), 
\quad
\contraction[1ex]{}{\psi}{_f^T(x)}{\bar{\psi} }
\psi_f^T(x) \bar{\psi}_f^T(y) = (K\gamma_5)  Q^{-1}(-\mu) \Delta^\dag(-\mu)  (K\gamma_5), \nonumber \\
\contraction[1ex]{}{\psi}{_2(x)}{\psi}
\psi_2(x) \psi_1^T(y) &=& J Q^{-1}(\mu) K,
\quad
\contraction[1ex]{}{\bar{\psi}}{_2(x)}{\bar{\psi}}
\bar{\psi}_2^T(x) \bar{\psi}_1(y) =J  (K\gamma_5)  Q^{-1}(-\mu) \gamma_5 ,
\eeq
with the flavor index $f=1,2$. 
Here $Q(\mu)=\Delta^\dag(\mu) \Delta(\mu) + J^2$. 
The upper-left equation shows the normal propagator, and the upper-right one is associated with the backpropagation.
The lower-left and right ones are the so-called anomalous propagators, which represent a quark to antiquark and an antiquark to quark change, respectively. 
The lower propagators come from the diquark source term in the action, as is evident from the fact that they are proportional to $J$.

\subsection{2-point correlation functions}
The calculation of the $2$-point correlation function in dense $2$-color QCD has been performed by Hands et al.~\cite{Hands2007-vp}, and we follow a line of their definitions.
The operators for the pion and rho meson ($I=1, J^P=0^-$ and $J^P=1^-$) are given by $M^1 = \bar{\psi}_1 \Gamma \psi_2$ or $\bar{\psi}_2 \Gamma \psi_1$, where $\Gamma=\gamma_{5}$ for the pion and $\Gamma=\gamma_1$ for the rho meson. 
The 2-point correlation functions read
\beq
\label{eq:2ptfunc_pirho}
\langle M^1(t,\vec{x}) M^{1 \dag}(0, \vec{y}) \rangle 
&=& 
\Tr[S_N(t,\vec{x}|0,\vec{y}) \bar{\Gamma} 
S_N(0,\vec{y}|t,\vec{x})\Gamma]
\nonumber
\\
&-& \Tr[S_A(t,\vec{x}|0,\vec{y}) \bar{\Gamma}^T
\bar{S}_A(0,\vec{y}|t,\vec{x})\Gamma].
\eeq
Here $\Tr[\cdot]$ denotes the trace in the color and spinor spaces and $\bar{\Gamma} = \gamma_4 \Gamma^\dag \gamma_4$. The first and second terms of the right side represent the contributions of the normal and anomalous propagators, respectively. 

The meson operators with $I=0$, $J^P=0^{\pm}$ are defined by $M^0 = (\bar{\psi}_1 \Gamma \psi_1 +  \bar{\psi}_2 \Gamma \psi_2)/\sqrt{2}$, where $\Gamma=1$ for $0^+$ and $\Gamma=\gamma_5$ for $0^-$, and then their 2-point correlation functions are given by
\beq
\label{eq:2ptfunc_I0meson}
\langle M^0(t,\vec{x}) M^{0 \dag}(0, \vec{y}) \rangle = 
&-&2 \Tr[S_N(t,\vec{x}|t,\vec{x}) \Gamma]
\Tr[S_N(0,\vec{y}|0,\vec{y})\bar{\Gamma}] \nonumber
\\
&+& \Tr[S_N(t,\vec{x}|0,\vec{y}) \bar{\Gamma}
S_N(0,\vec{y}|t,\vec{x})\Gamma] \nonumber
\\
&+& \Tr[S_A(t,\vec{x}|0,\vec{y}) \bar{\Gamma}^T
\bar{S}_A(0,\vec{y}|t,\vec{x})\Gamma].
\eeq
The first term of the right side corresponds to the disconnected diagram. 
Note that the third term has the opposite sign to that of the second term in Eq.~(\ref{eq:2ptfunc_pirho}).

The diquark operators in $I=0, J^P=0^{\pm}$ channel are defined by $D^0 =( \psi^T_1 K \bar{\Gamma} \psi_2 -  \psi^T_2 K \bar{\Gamma} \psi_1)/\sqrt{2}$, leading to the 2-point correlation functions:
\beq
\label{eq:2ptfunc_I0diquark}
\langle D^0(t,\vec{x}) D^{0 \dag}(0, \vec{y}) \rangle 
=
&-&2\Tr[\bar{S}_A(t,\vec{x}|t,\vec{x}) \Gamma K]
\Tr[S_A(0,\vec{y}|0,\vec{y}) K \bar{\Gamma}] \nonumber
\\
&-& \Tr[S_N(t,\vec{x}|0,\vec{y}) \Gamma K 
\bar{S}_N(0,\vec{y}|t,\vec{x})K\bar{\Gamma}] \nonumber
\\
&-& \Tr[S_N(t,\vec{x}|0,\vec{y}) K \Gamma^T
\bar{S}_N(0,\vec{y}|t,\vec{x})K\bar{\Gamma}].
\eeq
The first term of the right side corresponds to the disconnected diagram composed of the anomalous propagators, while the second and third terms are the contributions from the normal propagators.

\section{Simulation details}
In our simulation, we set $\beta=0.8$ and $\kappa=0.159$ on $32^4$ lattices, corresponding to the ratio $m^{0}_{\pi}/m^{0}_{\rho} \approx 0.81$ ($m^{0}_{\rho}$: the rho meson mass at $\mu=0$), lattice spacing $a \approx 0.16~\textrm{fm}$, and the temperature $T \approx 0.19 T_c$ with the pseudo-critical temperature of the chiral phase transition $T_c$ assumed to be $T_c=200 ~\textrm{MeV}$~\cite{Iida:2020emi}. In this setup, $m^{0}_{\pi}\approx 738~\textrm{MeV}$.
We use 400 configurations, and statistical uncertainties are estimated by the jackknife analysis at the bin size 40. 
We take periodic boundary conditions in the spatial directions while anti-periodic boundary conditions in the time direction. 
We use wall-type quark operators at the source and local quark operators at the sink. 

According to Ref.~\cite{Iida:2019rah}, there is a phase transition between the hadronic and superfluid phases around $\mu/m^{0}_{\pi}=0.5$. 
In the present simulation,  we take $j=0$ for sea quarks and $j=0.001$ for valence quarks in the range of  $0 \leq \mu/m^{0}_{\pi} \leq 0.40$, while we take $j=0.02$ for both sea and valence quark for $\mu/m^{0}_{\pi} >0.40$.
The diquark source term in the action is added to break the U($1$)$_B$ symmetry explicitly; by taking  the $j \to 0$ limit we can compare numerical data with predictions from several analytical studies.
The extrapolation $j \to 0$ is left for future works.

In this paper, we show the results that have the disconnected diagrams subtracted out in Eq.~(\ref{eq:2ptfunc_I0meson}) and Eq.~(\ref{eq:2ptfunc_I0diquark}). 
We found that the contributions coming from the disconnected diagrams do not change the mass spectrum drastically for almost all the hadrons in a small $\mu$ regime. 
The meson in $I=0$, $0^{+}$ channel is the exception, in which the disconnected diagrams give rise to large fluctuations since this channel can be coupled with the QCD vacua.
Further consideration of their contributions would be required in the future.

\section{Results}

\subsection{pion and rho meson ($I=1, J^P=0^-$ and $J^P=1^-$ channels)}
We first investigate the pion and rho meson spectrum, which are well-known from previous works~\cite{Muroya2002-qc, Hands2007-vp,Wilhelm:2019fvp}.
\begin{figure}[t]
    \centering
            \includegraphics[width=0.49\textwidth]{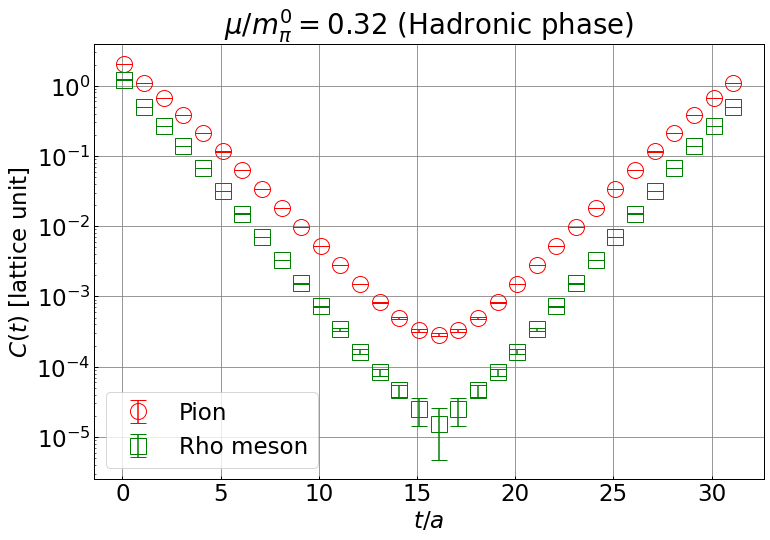}
	    \includegraphics[width=0.48\textwidth]{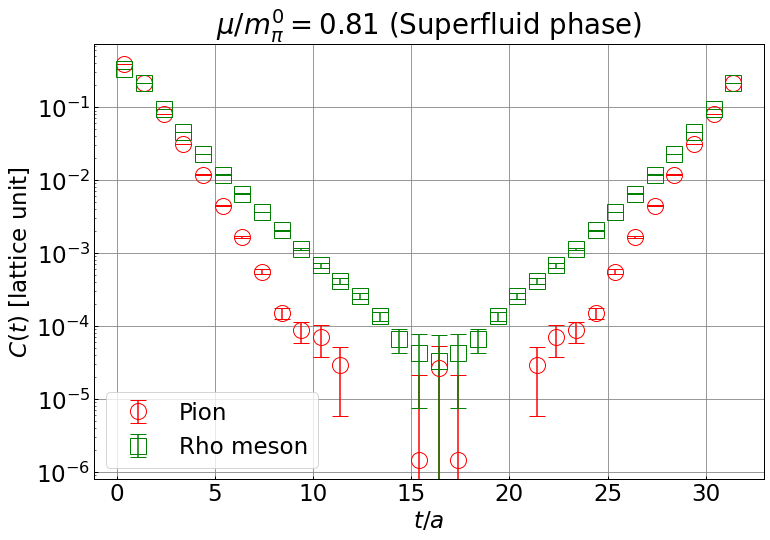}
	    \caption{ $2$-point correlation functions for the pion (red circles) and the rho meson (green squares) at $\mu/m^{0}_{\pi} = 0.32$ (left panel) and at $\mu/m^{0}_{\pi} = 0.81$ (right panel).}
	    \label{fig:pirho_2pt}
\end{figure}

In Fig.~\ref{fig:pirho_2pt}, we show the typical plots of the $2$-point correlation functions for the pion and rho meson. Here we take $\mu/m^{0}_{\pi} = 0.32$ as a typical case of the hadronic phase in the left panel, and  $\mu/m^{0}_{\pi} = 0.81$ as a typical case of the superfluid phase in the right panel.  
At $\mu/m^{0}_{\pi} = 0.32$, both signals are clear and the pion 2-point correlation function is always larger than  the rho meson at every timeslice.
In fact, the inequality, $C_{\pi} (t) \geq C_{\rho}(t)$, is proven analytically at zero chemical potential. Although the inequality requires the $\gamma_5$ hermiticity, our result indicates that the inequality is valid even in a small $\mu$ regime.
On the other hand, at $\mu/m^{0}_{\pi} = 0.81$, our results obviously break the inequality and the slope of the pion $2$-point function is much steeper than that of the rho meson, which indicates that pion becomes heavier than the rho meson in the superfluid phase.
We also found that the pion $2$-point function gets noisier at larger $\mu$.
It suggests that a $2$-body decay channel is beginning to open up. 

\begin{figure}[t]
    \centering
            \includegraphics[width=0.6\textwidth]{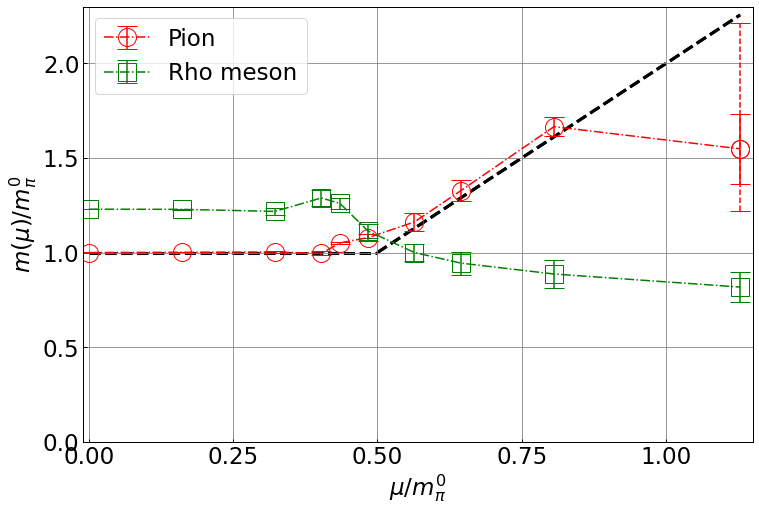}
	    \caption{Pion (red circles) and rho meson (green squares) masses at each chemical potential $\mu$. Solid and dashed errorbars correspond to the statistical errors and the statistical and systematic errors added in quadrature, respectively. Black dashed line shows the results for the pion mass in the chiral perturbation analysis~\cite{Kogut:1999iv, Kogut2000-so}.}
	    \label{fig:pirho_mass}
\end{figure}

Now, we summarize the $\mu$ dependence of the pion and rho meson masses as shown in Fig.~\ref{fig:pirho_mass}~\footnote{
Here, the effective masses are estimated by fitting the $2$-point functions using the cosh function.
The solid error bars denote the statistical errors. We also show the systematic error of the pion mass at $\mu/m^{0}_{\pi}=1.13$ as a dashed error bar.
It comes from the fit range dependence. Only this data point has a sizable systematic error.}.
At $\mu/m^{0}_{\pi} \leq 0.32$, both the pion and rho meson masses are independent of $\mu$, keeping the rho meson heavier than the pion. 
For $\mu/m^{0}_{\pi} \geq 0.40$, the pion mass starts to increase while the rho meson mass slightly increases and then starts to decrease.
Finally, the ordering of the pion and rho meson flips around $\mu/m^{0}_{\pi} = 0.5$, which is the transition point between the hadronic phase and the superfluid phase.
For $\mu/m^{0}_{\pi}>0.5$, the pion mass increases linearly while the rho meson mass decreases monotonically.

The results in small-$\mu$ regimes are consistent with Eq.~\eqref{eq:hadron-linear-mass}.
Here, $Q_{B}=0$ for mesons, $Q_{B}=2$ for diquarks, and $Q_{B}=-2$ for antidiquarks. 
Furthermore, the linear increase of the pion mass in the superfluid phase is also almost consistent with the chiral model such as the chiral perturbation theory (ChPT)~\cite{Kogut:1999iv, Kogut2000-so}, which is shown as a black dashed line in the figure. 
We see slight deviations between our data and the result of ChPT only around $\mu/m^{0}_{\pi} = 0.5$.
It is due possibly to an effect of the finite diquark source term.
Furthermore, the decreasing behavior of the rho meson mass in a high-$\mu$ regime has been predicted in the analytical studies for dense $3$-color QCD~\cite{Brown:1991kk,Hatsuda:1991ez}. 
Our results in the superfluid phase seem consistent with such arguments in some way or another.

\subsection{Hadrons in $I=0$, $J^{P}=0^{\pm}$ channels}
\subsubsection{2-point functions}
\begin{figure}[t]
    \centering
            \includegraphics[width=0.49\textwidth]{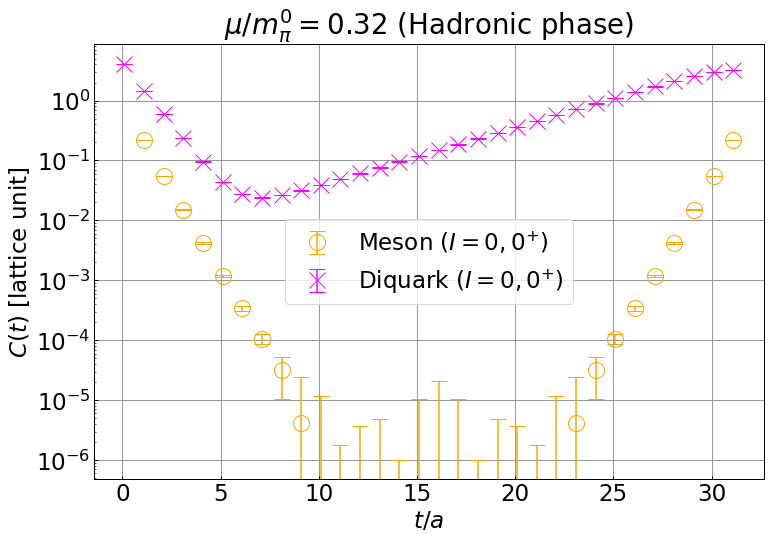}
	    \includegraphics[width=0.48\textwidth]{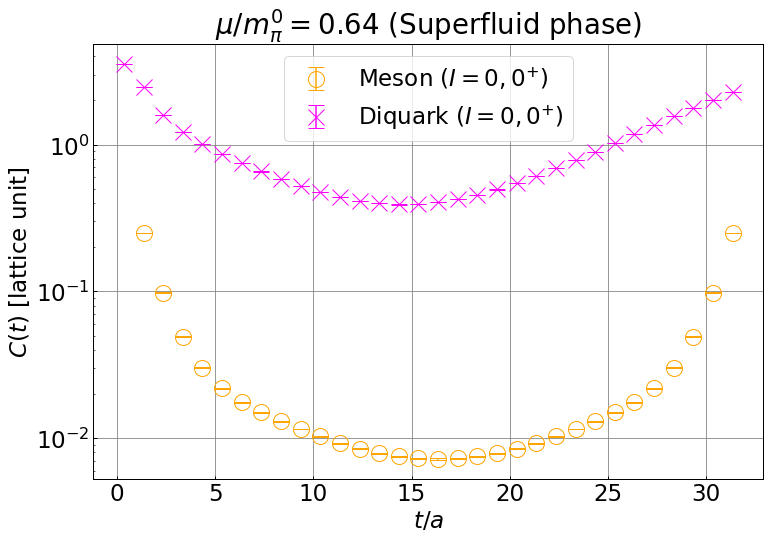}
	    \caption{Meson (orange circles) and diquark (magenta crosses) 2-point correlation functions in isospin $I=0$ and $0^{+}$ channel at $\mu/m^{0}_{\pi} = 0.32$ (left) and at $\mu/m^{0}_{\pi} = 0.64$ (right).}
	    \label{fig:scal_2pt}
\end{figure}
Figure~\ref{fig:scal_2pt} presents the 2-point functions of the meson and diquark in isospin $I=0$ and $J^{P}=0^{+}$ channel at $\mu/m^{0}_{\pi} = 0.32$ and at $\mu/m^{0}_{\pi} = 0.64$. 
At $\mu/m^{0}_{\pi} = 0.32$ as a typical case of the hadronic phase, the meson 2-point function is symmetric under time reversal, namely, $t/a \leftrightarrow (N_t -t/a)$, while the diquark one is asymmetric and the slope is steeper in the forward direction than that in the backward direction. 
On the other hand, at $\mu/m^{0}_{\pi} = 0.64$ as a typical case of the superfluid phase, the signals of both 2-point functions are clear and the slopes are gradual, which indicates that both meson and diquark masses decrease in the superfluid phase. 
Furthermore, the diquark 2-point function becomes symmetric in the superfluid phase.

\begin{figure}[t]
    \centering
            \includegraphics[width=0.49\textwidth]{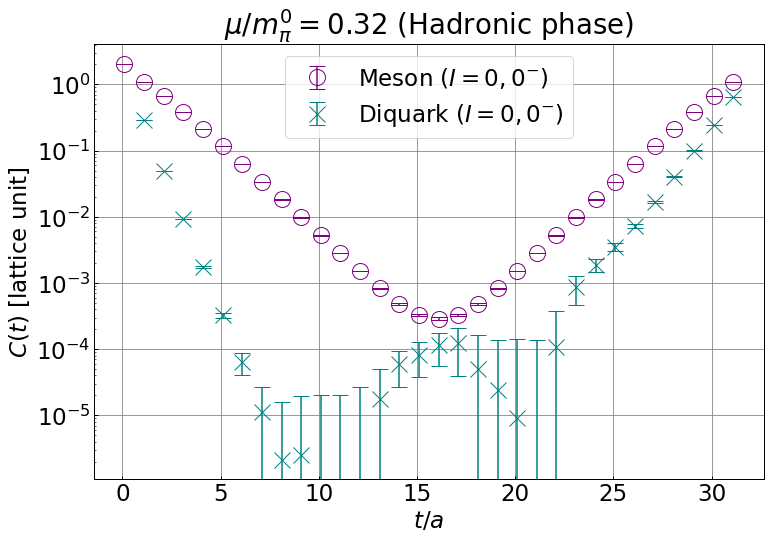}
	    \includegraphics[width=0.48\textwidth]{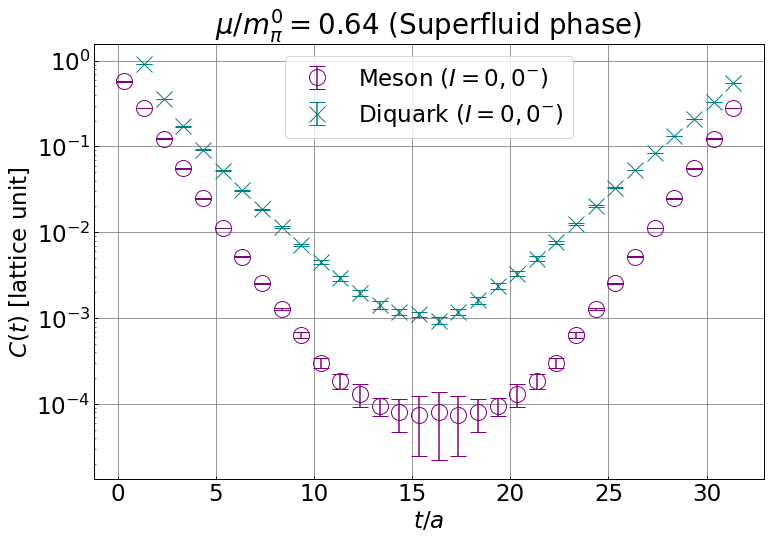}
	    \caption{Meson (purple circles) and diquark (cyan crosses) 2-point correlation functions in isospin $I=0$ and $0^{-}$ channel at $\mu/m^{0}_{\pi} = 0.32$ (left) and at $\mu/m^{0}_{\pi} = 0.64$ (right).}
	    \label{fig:pscal_2pt}
\end{figure}

The 2-point functions in $I=0, J^P=0^{-}$ channel are shown in Fig.~\ref{fig:pscal_2pt}.
As to time reversibility, they are similar to those in the $I=0, J^P=0^{+}$ channel; the meson 2-point function is symmetric and the diquark one is asymmetric at $\mu/m^{0}_{\pi} = 0.32$, while both are symmetric at $\mu/m^{0}_{\pi} = 0.64$.  
However, the diquark 2-point function has a steeper slope than the meson one at $\mu/m^{0}_{\pi} = 0.32$ in $I=0, J^P=0^{-}$ channel.
Furthermore, in contrast to the $I=0, J^P=0^{+}$ channel, the meson 2-point function at $\mu/m^{0}_{\pi} = 0.64$ gets noisy compared with that at $\mu/m^{0}_{\pi} = 0.32$, and the slopes in early timeslices ($t/a \simeq 4$--$7$) and in late timeslices ($t/a \simeq 9$--$12$) look different from each other. 
The behavior suggests that there is a comparable contribution from the first-excited state.

\subsubsection{Mass spectrum}
Now, we extract the effective masses from the corresponding $2$-point functions.
If the $2$-point function is symmetric under $t/a \leftrightarrow (N_t -t/a)$ theoretically, as is the case with all the $2$-point functions at $\mu =0$ and the meson $2$-point functions in all $\mu$ regimes, then we fit it using a single cosh function.
As for the diquark (antidiquark) $2$-point functions at $\mu \ne 0$, which are asymmetric, we utilize a single-exponential function to fit each $2$-point function in the backward and forward directions separately and define the diquark and antidiquark masses by the fit results in the backward and forward directions, respectively.
Furthermore, for the $2$-point functions that are expected to have a large contribution from the first-excited state, we use the double-cosh function $C_{0}\cosh(m_{0}(t-T/2))+C_{1}\cosh(m_{1}(t-T/2))$ in the fitting and pick up the ground-state masses.
We apply it to the $I=0$, $0^{+}$ meson at $\mu/m^{0}_{\pi}=1.13$ and to the $I=0$, $0^{-}$ meson at $\mu/m^{0}_{\pi}=0.64$, $0.81$, and $1.13$.

\begin{figure}[t]
    \centering
            \includegraphics[width=0.49\textwidth]{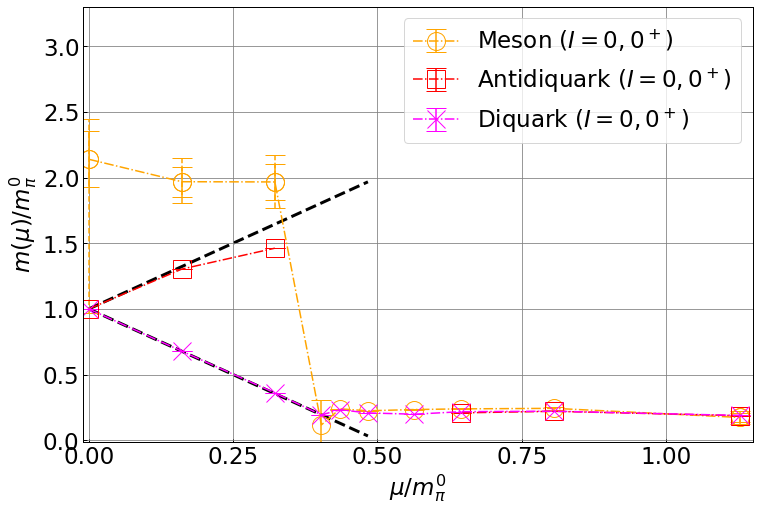}
	    \includegraphics[width=0.48\textwidth]              {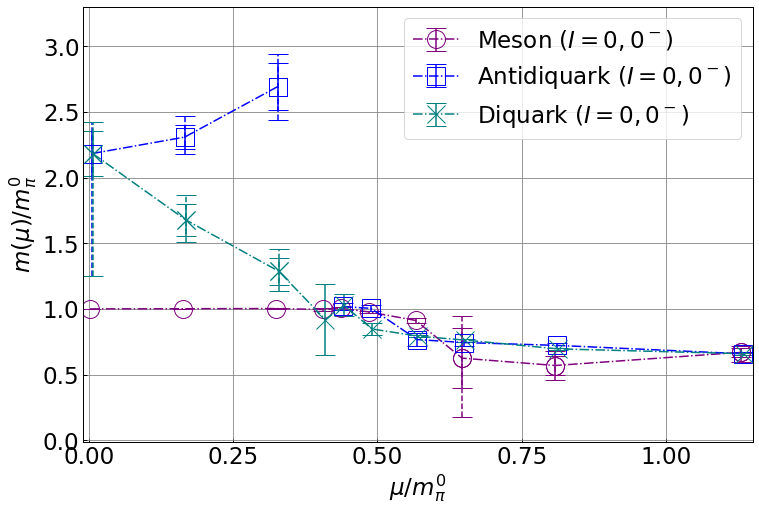}
	    \caption{Hadron masses at each $\mu$ in $I=0$, $0^{+}$ channel (left) and $I=0$, $0^{-}$ channel (right). Solid errorbars correspond to the statistical errors, and dashed errorbars correspond to the statistical and systematic errors added in quadrature. Black dashed line shows the results for the diquark and antidiquark masses in the $I=0$, $0^{+}$ channel in the chiral perturbation analysis~\cite{Kogut:1999iv, Kogut2000-so}.}
	    \label{fig:scalpscal_mass}
\end{figure}

The results for hadron masses in $I=0$, $0^{\pm}$ channel at each $\mu$ are presented in Fig.~\ref{fig:scalpscal_mass}~\footnote{For the $I=0$, $0^{+}$ antidiquark at $\mu/m^{0}_{\pi}=0.40$--$0.56$ and the $I=0$, $0^{-}$ antidiquark at $\mu/m^{0}_{\pi}=0.40$, the results are not presented because the effect of the anti-periodicity in the 2-point functions is too large to extract their masses.}. 
At zero chemical potential, the masses of the diquark and antidiquark in $I=0$, $0^{+}$ channel are the same as the pion mass, which is guaranteed by the unbroken part of the Pauli-Gursey symmetry. 
The $I=0$, $0^{-}$ meson, which is associated with the eta meson in three-color QCD, also has the same mass as that of the pion. 
The reason for the eta-pion degeneracy is that we neglect the disconnected diagrams, which are associated with the $U(1)_{A}$ anomaly.
Compared with these hadrons, the $I=0$, $0^{+}$ meson and the $I=0$, $0^{-}$ diquark/antidiquark are much heavier. 

At $\mu/m^{0}_{\pi} \ll 1$ in the hadronic phase, the hadron masses behave as Eq.~\eqref{eq:hadron-linear-mass} in each channel.
For $I=0$, $0^{+}$ channel, this is consistent with the results of ChPT~\cite{Kogut:1999iv, Kogut2000-so}.
However, we see the apparent discrepancy in the $I=0$, $0^{+}$ antidiquark at $\mu/m^{0}_{\pi} = 0.32$.
Here, we take small $t/a$ data since the region of the forward propagation for the diquark $2$-point function is narrow. The result for the mass suffers from the excited states in such a $t/a$ regime.

In the superfluid phase ($\mu/m^{0}_{\pi} \geq 0.5$), the three hadron masses in each $I=0$, $0^{\pm}$ channel are degenerate. It indicates the mixing among the three hadrons in the ground state due to the spontaneous breaking of $U(1)_{B}$ symmetry. 
In particular, the masses of the $I=0$, $0^{+}$ hadrons are very small and steady against change in $\mu$, indicating that the $2$-point functions couple to the (pseudo) Nambu-Goldstone mode associated with the diquark condensation. 
In our understanding, the mass suffers from the effect of the diquark source term in the action. It is expected to be massless if we take the $j=0$ limit. 
For the $I=0$, $0^{-}$ hadrons, the masses get small but larger than the $I=0$, $0^{+}$ hadron masses, and decrease when $\mu$ increases. 

Such a mixing phenomenon between mesons and diquarks in the superfluid phase appears also in the linear-sigma model based on the approximate $SU(4)$ Pauli-Gursey symmetry with the chiral and diquark mean fields~\cite{Suenaga:2022uqn}. 
Furthermore, the hadron spectrum in these channels is almost consistent with our numerical results.
These consistencies support our interpretation of the numerical results, that is, the meson-diquark-antidiquark mixing occurs in the superfluid phase.

\section{Summary and discussion}
We study the hadron spectrum in 2-color QCD in a low-temperature and finite-density regime where both the hadronic phase and the superfluid phase appear. 
To avoid the numerical instability, we use the fermion action with the diquark source term.
It induces the anomalous propagators, which give rise to a quark to antiquark and an antiquark to quark change.

The pion and rho meson masses in our results are constant at small $\mu$, which is consistent with the theoretical prediction that the hadron masses behave as Eq.~\eqref{eq:hadron-linear-mass} at small $\mu$. 
These masses begin to change before the hadronic-superfluid phase transition. 
Accordingly, the spectral ordering flips around the phase transition point; the pion mass increases linearly while the rho meson mass monotonically decreases. 
The behavior is consistent with the previous lattice studies and with several analytical predictions. 

The hadrons masses in $I=0, J^P=0^{\pm}$ channels at small $\mu$ are consistent with Eq.\ (\ref{eq:hadron-linear-mass})
In the superfluid phase, all the hadrons in each channel have almost the same mass. 
This indicates that meson-diquark-antidiquark mixing happens due to the breaking of $U(1)_{B}$ symmetry in the superfluid phase, which is also seen in the linear sigma model based on the approximate $SU(4)$ Pauli-Gursey symmetry with the chiral and diquark mean fields. 

In this work, we neglect the disconnected diagrams, which are directly related to the effect of the $U(1)_{A}$ anomaly, which breaks the degeneracy of the spectrum of the chiral partner. 
Examination of such contribution is of interest in that the magnitude of the anomaly effect determines the asymptotic behavior of the eta meson mass at infinitely large $\mu$, as seen in the linear sigma model~\cite{Suenaga:2022uqn}.
Furthrmore, in this work, we use finite $j$ at large $\mu$. Taking the $j \to 0$ extrapolation is necessary in order to obtain clean results in the superfluid phase, for example, the $I=0$, $0^{+}$ masses that are expected to be zero in this limit. 
These corrections should be considered in the future.

\begin{acknowledgments}
We would like to thank S.~Aoki for useful conversations.
K.~M. is supported in part by JST SPRING, Grant Number JPMJSP2110, and by the Japan Society for the Promotion of Science (JSPS).
The numerical simulation is supported by the HPCI-JHPCN System Research Project (Project ID: jh220021). D.~S. is supported by the RIKEN special postdoctoral researcher program.
The work of E.~I. is supported by JSPS KAKENHI with Grant Number 19K03875, JST PRESTO Grant Number JPMJPR2113, JSPS Grant-in-Aid for Transformative Research Areas (A) JP21H05190 and JST Grant Number JPMJPF2221, and the work of K.~I. is supported by JSPS KAKENHI with Grant Number 18H05406.

\end{acknowledgments}
\bibliographystyle{utphys}
\bibliography{2color_spectrum}

\providecommand{\href}[2]{#2}\begingroup\raggedright\begin{thebibliography}{10}

\bibitem{Brown:1991kk}
G.~E. Brown and M.~Rho, ``{Scaling effective Lagrangians in a dense medium},''
  \href{http://dx.doi.org/10.1103/PhysRevLett.66.2720}{{\em Phys. Rev. Lett.}
  {\bfseries 66} (1991) 2720--2723}.

\bibitem{Hatsuda:1991ez}
T.~Hatsuda and S.~H. Lee, ``{QCD sum rules for vector mesons in the nuclear
  medium},'' \href{http://dx.doi.org/10.1103/PhysRevC.46.R34}{{\em Phys. Rev.
  C} {\bfseries 46} no.~1, (1992) R34}.

\bibitem{Murakami}
K.~Murakami, K.~Iida, and E.~Itou. in preparation.

\bibitem{Iida:2022hyy}
K.~Iida and E.~Itou, ``{Velocity of Sound beyond the High-Density Relativistic
  Limit from Lattice Simulation of Dense Two-Color QCD},''
  \href{http://arxiv.org/abs/2207.01253}{{\ttfamily arXiv:2207.01253
  [hep-ph]}}.

\bibitem{Muroya2002-qc}
S.~Muroya, A.~Nakamura, and C.~Nonaka, ``Behavior of hadrons at finite density
  -- lattice study of color {SU(2}) {QCD},''
  \href{http://dx.doi.org/10.1016/S0370-2693(02)03065-4}{{\em Phys. Lett. B}
  {\bfseries 551} (2003) 305--310},
  \href{http://arxiv.org/abs/hep-lat/0211010}{{\ttfamily
  arXiv:hep-lat/0211010}}.

\bibitem{Hands2007-vp}
S.~Hands, P.~Sitch, and J.-I. Skullerud, ``Hadron spectrum in a {Two-Colour}
  {Baryon-Rich} medium,''
  \href{http://dx.doi.org/10.1016/j.physletb.2008.01.078}{{\em Phys. Lett. B}
  {\bfseries 662} (2008) 405--412},
  \href{http://arxiv.org/abs/0710.1966}{{\ttfamily arXiv:0710.1966 [hep-lat]}}.

\bibitem{Wilhelm:2019fvp}
J.~Wilhelm, L.~Holicki, D.~Smith, B.~Wellegehausen, and L.~von Smekal,
  ``{Continuum Goldstone spectrum of two-color QCD at finite density with
  staggered quarks},''
  \href{http://dx.doi.org/10.1103/PhysRevD.100.114507}{{\em Phys. Rev. D}
  {\bfseries 100} no.~11, (2019) 114507},
  \href{http://arxiv.org/abs/1910.04495}{{\ttfamily arXiv:1910.04495
  [hep-lat]}}.

\bibitem{Iida:2019rah}
K.~Iida, E.~Itou, and T.-G. Lee, ``Two-colour {QCD} phases and the topology at
  low temperature and high density,''
  \href{http://dx.doi.org/10.1007/JHEP01(2020)181}{{\em JHEP} {\bfseries 01}
  (2020) 181}, \href{http://arxiv.org/abs/1910.07872}{{\ttfamily
  arXiv:1910.07872 [hep-lat]}}.

\bibitem{Iida:2020emi}
K.~Iida, E.~Itou, and T.-G. Lee, ``{Relative scale setting for two-color QCD
  with $N_f$=2 Wilson fermions},''
  \href{http://dx.doi.org/10.1093/ptep/ptaa170}{{\em PTEP} {\bfseries 2021}
  no.~1, (2021) 013B05}, \href{http://arxiv.org/abs/2008.06322}{{\ttfamily
  arXiv:2008.06322 [hep-lat]}}.

\bibitem{Kogut:1999iv}
J.~B. Kogut, M.~A. Stephanov, and D.~Toublan, ``{On two color QCD with baryon
  chemical potential},''
  \href{http://dx.doi.org/10.1016/S0370-2693(99)00971-5}{{\em Phys. Lett. B}
  {\bfseries 464} (1999) 183--191},
  \href{http://arxiv.org/abs/hep-ph/9906346}{{\ttfamily arXiv:hep-ph/9906346}}.

\bibitem{Kogut2000-so}
J.~B. Kogut, M.~A. Stephanov, D.~Toublan, J.~J.~M. Verbaarschot, and
  A.~Zhitnitsky, ``{QCD-like} theories at finite baryon density,''
  \href{http://dx.doi.org/10.1016/S0550-3213(00)00242-X}{{\em Nucl. Phys. B}
  {\bfseries 582} (2000) 477--513},
  \href{http://arxiv.org/abs/hep-ph/0001171}{{\ttfamily arXiv:hep-ph/0001171}}.

\bibitem{Suenaga:2022uqn}
D.~Suenaga, K.~Murakami, E.~Itou, and K.~Iida, ``{Probing hadron mass spectrum
  in dense two-color QCD with linear sigma model},''
  \href{http://arxiv.org/abs/2211.01789}{{\ttfamily arXiv:2211.01789
  [hep-ph]}}.

\end{thebibliography}\endgroup

\end{document}